\documentclass[aps,prd,showpacs,nofootinbib,twocolumn,10pt]{revtex4-1}
\usepackage{color,amsmath,amssymb,graphicx,latexsym,lineno}
\usepackage{threeparttable,multirow,txfonts,subfigure,booktabs}

\def\aap{Astron.\ Astrophys.\ }
\def\aaps{Astron.\ Astrophys.\ Supp.\ }
\def\apj{Astrophys.\ J.\ }

\def\apjs{Astrophys.\ J.\ Supp.\ }
\def\aj{Astron.\ J.\ }
\def\mnras{Mon.\ Not.\ Roy.\ Astron.\ Soc.\ }

\def\prd{Phys.\ Rev.\ D\ }
\def\prl{Phys.\ Rev.\ Lett.\ }

\begin{document}
%\linenumbers
%setpagewiselinenumbers
%modulolinenumbers[5]
%linenumbers
%\modulolinenumbers[5]
%pagewiselinenumbers
%\switchlinenumbers

\title{Extended Very-High-Energy Gamma-ray Emission Surrounding PSR J0622 + 3749 Observed by LHAASO-KM2A}

\author{
F. Aharonian$^{26,27}$,
Q. An$^{4,5}$,
Axikegu$^{20}$,
L.X. Bai$^{21}$,
Y.X. Bai$^{1,3}$,
Y.W. Bao$^{15}$,
D. Bastieri$^{10}$,
X.J. Bi$^{1,2,3}$,
\\
Y.J. Bi$^{1,3}$,
H. Cai$^{23}$
% ...
}
\address{%
$^1$Key Laboratory of Particle Astrophyics \& Experimental Physics Division \& Computing Center, Institute of High Energy Physics, Chinese Academy of Sciences, 100049 Beijing, China\\
$^2$University of Chinese Academy of Sciences, 100049 Beijing, China\\
$^3$TIANFU Cosmic Ray Research Center, Chengdu, Sichuan,  China\\
$^4$State Key Laboratory of Particle Detection and Electronics, China\\
$^5$University of Science and Technology of China, 230026 Hefei, Anhui, China\\
% ...
$^{10}$Center for Astrophysics, Guangzhou University, 510006 Guangzhou, Guangdong, China\\
% ...
$^{15}$School of Astronomy and Space Science, Nanjing University, 210023 Nanjing, Jiangsu, China\\
% ...
$^{20}$School of Physical Science and Technology \&  School of Information Science and Technology, Southwest Jiaotong University, 610031 Chengdu, Sichuan, China\\
$^{21}$College of Physics, Sichuan University, 610065 Chengdu, Sichuan, China\\
% ...
$^{23}$School of Physics and Technology, Wuhan University, 430072 Wuhan, Hubei, China\\
% ...
$^{26}$Dublin Institute for Advanced Studies, 31 Fitzwilliam Place, 2 Dublin, Ireland \\
$^{27}$Max-Planck-Institut for Nuclear Physics, P.O. Box 103980, 69029  Heidelberg, Germany \\
% ...
}
%\collaboration{LHAASO Collaboration}
\date{\today}
\noaffiliation
\begin{abstract}
We report the discovery of an extended very-high-energy (VHE) gamma-ray 
source around the location of the middle-aged (207.8 kyr) pulsar 
PSR J0622+3749 with the Large High Altitude Air Shower Observatory (LHAASO). 
The source is detected with a significance of $8.2\sigma$ for $E>25$~TeV 
assuming a Gaussian template. The best-fit location is
(R.A., Dec.)$=(95^{\circ}\!.47\pm0^{\circ}\!.11,\,37^{\circ}\!.92
\pm0^{\circ}\!.09)$, and the extension is $0^{\circ}\!.40\pm0^{\circ}\!.07$.
The energy spectrum can be described by a power-law spectrum with an
index of ${-2.92 \pm 0.17_{\rm stat} \pm 0.02_{\rm sys} }$.
No clear extended multi-wavelength counterpart of the LHAASO source
has been found from the radio to sub-TeV bands. The LHAASO observations
are consistent with the scenario that VHE electrons
escaped from the pulsar, diffused in the interstellar medium, and scattered
the interstellar radiation field. If interpreted as the pulsar
halo scenario, the diffusion coefficient, inferred for electrons with
median energies of $\sim160$~TeV, is consistent with those obtained
from the extended halos around Geminga and Monogem and much smaller
than that derived from cosmic ray secondaries.
The LHAASO discovery of this source thus likely enriches the class of
so-called pulsar halos and confirms that high-energy particles
generally diffuse very slowly in the disturbed medium around pulsars.
\end{abstract}


\maketitle

\section{Introduction}

Charged cosmic rays (CRs) are known to propagate diffusively in the
random magnetic field of the Milky Way. The diffusion coefficient,
which relies on the properties of the turbulent interstellar medium
(ISM), is a key parameter governing the propagation of CRs.
Through measuring the secondary-to-primary ratios of CR nuclei,
the average diffusion coefficient of the Milky Way can be inferred
\cite{2011ApJ...729..106T,2017PhRvD..95h3007Y}.
While the propagation of CRs is naturally expected to be
inhomogeneous, the simple uniform and isotropic diffusion model
can account for most of the measurements of CRs and diffuse
$\gamma$ rays \cite{2007ARNPS..57..285S}.

Recently, the HAWC collaboration reported the observations of
extended very-high-energy (VHE) $\gamma$-ray emission from two
middle-aged, isolated pulsars, Geminga and Monogem
\cite{2017Sci...358..911A}. The spatial morphologies of the
$\gamma$-ray emission indicate that the diffusion of particles
in the regions around those pulsars are much slower than the
average value to give enough secondary particle yields
\cite{2016PhRvL.117w1102A,2017PhRvD..95h3007Y}.
These results suggest that the diffusion of particles in the
Milky Way is very likely inhomogeneous \cite{2018ApJ...863...30F, 2018PhRvD..97l3008P,2018PhRvD..98h3009H,2019MNRAS.484.3491T}.
Such extended halos around middle-aged pulsars\footnote{Note that
different definitions of pulsar halos exist in the literature
\cite{2019PhRvD.100d3016S,2020A&A...636A.113G}. For example,
some young pulsars that exhibit particle escape are also called
pulsar halos \cite{2019ApJ...877...54B,2019ApJ...881..148B}.}
may be common at VHE energies \cite{2017PhRvD..96j3016L,
2020PhRvD.101j3035D,2020arXiv201205932D}, which was actually
predicted a long time ago \cite{2004vhec.book.....A}.
Pulsar halos may even contribute to the diffuse emission at TeV energies \cite{2000A&A...362..937A,2004vhec.book.....A,2018PhRvL.120l1101L}.
The cause of the slow diffusion is yet to be elucidated but might be
due to specific properties of the magnetic turbulence around the pulsar \cite{2018MNRAS.479.4526L,2019PhRvL.123v1103L,2019MNRAS.488.4074F}.

Here we report the detection of LHAASO J0621+3755, an extended 
$\gamma$-ray source with energies above 10 TeV, with half array of 
the LHAASO experiment. In the third HAWC source catalog, a source 
with similar coordinates, 3HWC J0621+382, was reported recently 
\cite{2020arXiv200708582A}. No source with similar coordinate 
was shown in the second HAWC source catalog \cite{2017ApJ...843...40A}.
However, the nearest Fermi source of 3HWC J0621+382 was found to be
the blazar 4FGL J0620.3+3804. LHAASO J0621+3755 is positionally
coincident with the $\gamma$-ray pulsar J0622+3749 discovered by
Fermi-LAT \cite{2012ApJ...744..105P}. The period of PSR J0622+3749
is about 0.333 s, the spin-down luminosity is $2.7\times10^{34}$
erg~s$^{-1}$, and the characteristic age is about 207.8 kyr.
No precise distance measurement of the pulsar is available now.
A ``pseudo distance'' of 1.6 kpc was given via the correlation
between the $\gamma$-ray luminosity and the spin-down power for
$\gamma$-ray pulsars \cite{2010ApJ...725..571S}.
The VHE $\gamma$-ray and multi-wavelength properties of LHAASO
J0621+3755 have been studied, which suggest that this source is
very likely a pulsar halo similar to Geminga and Monogem
observed by HAWC \cite{2017Sci...358..911A}.

\section{LHAASO-KM2A observations}

\subsection{The LHAASO experiment}
LHAASO is a hybrid, large area, wide field-of-view observatory for
CRs and $\gamma$ rays in a wide energy range.
LHAASO serves as the most sensitive $\gamma$-ray detector for energies
above a few tens of TeV, and is expected to give revolutionary insights
in the VHE domain of astroparticle physics, such as the origin and
propagation of CRs, as well as the nature of VHE $\gamma$-ray sources.
KM2A is the main array of LHAASO, with an area of $\sim $1.3 km$^2$, consisting of 5195
electromagnetic detectors (EDs) and 1188 muon detectors (MDs). See the
{\tt Supplemental Material}~\cite{SM} for the detector configuration.

LHAASO's first observation on the Crab Nebula is presented in \cite{2020arXiv201006205A}.
By the measurements of this standard candle, detailed studies of the detector performance
have been carried out, including angular resolution, pointing accuracy and
cosmic-ray background rejection power.
The pipeline of data analysis and Monte-Carlo (MC) simulations was then constructed.
In this paper, we adopt the same simulation procedure and get the MC data sample
of $2.2\times10^8$ $\gamma$-ray
events as described in Ref.~\cite{2020arXiv201006205A}.

\subsection{Analysis method}
Data used in this analysis were collected by the half array of KM2A,
from December 27, 2019 to November 9, 2020, with a live time of 281.9 days.
The directions of $\gamma$ rays are reconstructed using the arrival
time and deposited energy recorded by each ED. The angular resolution
(68$\%$ containment) is $0^{\circ}\!.5-0^{\circ}\!.8$ at 20 TeV and
$0^{\circ}\!.24-0^{\circ}\!.30$ at 100 TeV, depending on the declination
of incident photons. The maximum zenith angle of events was chosen 
as 50$^{\circ}$. Event selection conditions are consistent with
those in the Crab Nebula analysis \cite{2020arXiv201006205A}, 
 and more details can be found in the {\tt Supplemental Material}~\cite{SM}.

KM2A uses $\rho_{50}$, defined as the particle
density in the best-fit Nishimura-Kamata-Greissen (NKG; \cite{1960Greisen})
function at a perpendicular distance of 50 m from the shower axis,
to estimate the primary energy of a $\gamma$-ray event.
This parameter is a robust energy estimator, because it utilizes the
whole knowledge of the lateral distribution function of a shower \cite{2017_kawata_energy,2019_abeysekara_measurement,2020arXiv201006205A}.
For showers with zenith angles less than 20$^{\circ}$, the energy resolution
is about 24$\%$ at 20 TeV and 13$\%$ at 100 TeV \cite{2020arXiv201006205A}.

Because a high-energy $\gamma$-ray-induced shower has fewer muons than a 
CR-induced shower, we use the ratio $N_{\mu}/N_e$ to discriminate
$\gamma$ rays from the CR background, where $N_{\mu}$ is the total number of
muons collected by MDs and $N_e$ is the total number of particles counted
by EDs. The criteria of this ratio were optimized using the MC
events of $\gamma$-ray photons and the real CR data. KM2A is capable of
rejecting the CR background by $99\%$ at 20 TeV and $99.99\%$ above 100 TeV,
while maintaining a $90\%$ efficiency for $\gamma$ rays \cite{2020arXiv201006205A}.

The sky around the target source is binned into cells with a size of
$0^{\circ}\!.1$ in both the right ascension (R.A.) and declination (Dec.)
directions. The background map is estimated by the equi-zenith angle
method \cite{2005amenomori_northern,2006amenomori_anisotropy}.
In brief, for a candidate source, the background is estimated by
collecting events in the same zenith angle belt, after excluding
events from the source region. The radius of the source region is
defined as three times the width of a Gaussian function, which is
the convolution of the point-spread function (PSF) and the source
extension. This method can eliminate various detection effects caused by
the instrumental and environmental variations.
Another widely used method is the so-called direct integration 
method~\cite{2020arXiv201006205A}, which estimates the background using 
the events in the same directions in horizontal coordinate but 
different arrival times. A cross check of the two methods showed no 
noticeable diﬀerence.

The significance of the source was estimated using a test statistic
variable as two times of the logarithmic likelihood ratio, i.e.,
$TS = 2\ln({\mathcal L}_{s+b}/{\mathcal L}_b)$, where ${\mathcal L}_{s+b}$
is the maximum likelihood for the signal plus background hypothesis and
${\mathcal L}_b$ is the likelihood for the background only hypothesis.
According to the Wilks' theorem~\cite{1938wilks}, in the background only
case, the TS value follows a $\chi^2$ distribution with $n$ degrees
of freedom, where $n$ is the number of free parameters in the signal model.
%The statistical significance of this test can be then obtained.
In the case of a point source with fixed position, which has only one free
parameter (the normalization when ignoring the spectral distribution),
the pre-trial significance is $\sqrt{TS}$.

We use a binned-likelihood, forward-folding procedure to measure the
spectral energy distribution (SED) of this source. The number of
$\gamma$-ray events is counted in three bins with a width of
$\Delta \log_{10}E=0.4$ (instead of 0.2 as in Ref.~\cite{2020arXiv201006205A}
to give a higher significance in each bin) ranging from 10 TeV to 160 TeV.
The SED of the source is assumed to follow a power-law spectrum
$dN/dE = \phi_0 \times \left(E/E_0\right)^{-\gamma}$,
where $E_0=40$~TeV is a reference energy. The best-fit values of
$\phi_0$ and $\gamma$ are obtained via the maximum likelihood algorithm.

\subsection{Results}

\begin{figure}[!htb]
\centering
\includegraphics[width=0.48\textwidth]{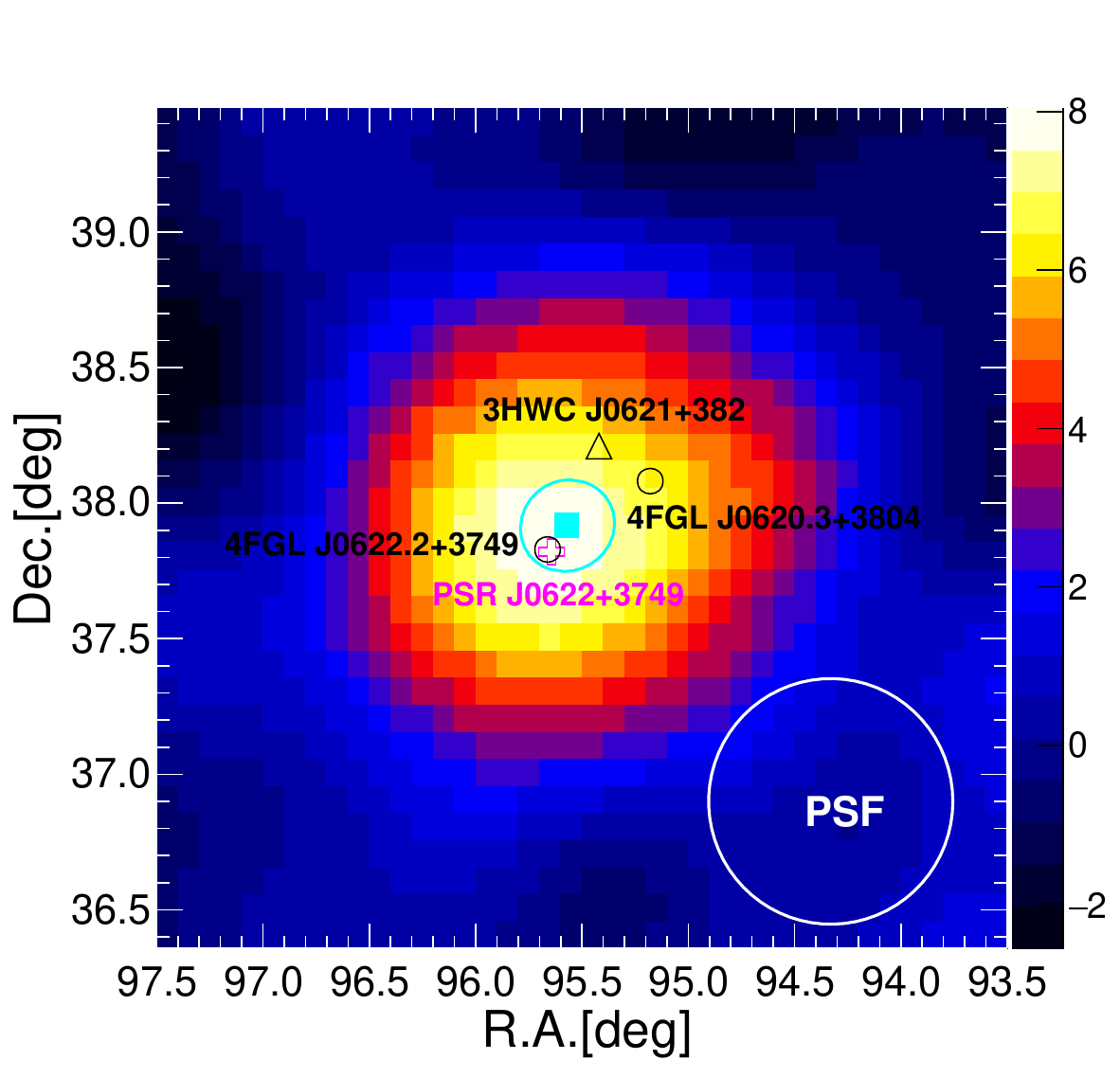}
\caption{Significance map of the $3^{\circ}\times3^{\circ}$ region
around LHAASO J0621+3755 with energy above 25 TeV. The cyan square and circle denote the best-fit
and $1\sigma$ range of the location of the LHAASO source. The triangle marks
the location of 3HWC J0621+382, the black circles show the locations of the
two 4FGL sources, and the pink cross marks the location of PSR J0622+3749.
The angular distance between the centroid of LHAASO J0621+3755 and PSR
J0622+3749 is $0^{\circ}\!.11 \pm 0^{\circ}\!.12$. The white circle at the
bottom-right corner shows the size of the LHAASO PSF ($68\%$ containment).}
\label{fig:map25TeV}
\end{figure}

Fig.~\ref{fig:map25TeV} shows the $3^{\circ}\times3^{\circ}$ significance map
around PSR J0622 + 3749 with energies above $25$~TeV in the equatorial coordinates.
%The point spread function (PSF) of KM2A can be approximately described by
%a two-dimensional Gaussian function.
This map is smoothed with the PSF, the 68\% containment radius of which
is $0^{\circ}\!.45$ in this energy range, as indicated by the white circle in Fig.1.

We use four spatial templates, convolved with the PSF, to study the morphology
of the source: point source, two-dimensional Gaussian model, uniform disk,
and the diffusion model from a point source with constant injection rate
\cite{2017Sci...358..911A}. Table~\ref{table:fit0} lists the best-fit source
positions, extensions, and the $TS$ values for these templates. For the point
source assumption, the fit to the $>25$~TeV skymap gives a $TS$ value of 63.0.

In the case of the two-dimensional Gaussian model,
the fit yields R.A.$=95^{\circ}\!.47 \pm 0^{\circ}\!.11$,
Dec.$=37^{\circ}\!.92 \pm 0^{\circ}\!.09$, and extension
$\sigma=0^{\circ}\!.40 \pm 0^{\circ}\!.07$.
The centroid of LHAASO J0621+3755 is consistent with the location
of Fermi-LAT pulsar J0622+3749, with in an angular distance of
$0^{\circ}\!.11 \pm 0^{\circ}\!.12$. It is also consistent with the expectation
of the pulsar halo model that the $\gamma$-ray emission above 10 TeV is
close to the pulsar due to the fast cooling of such VHE $e^{\pm}$,
even if there is a moderate proper motion of the pulsar
\cite{2020arXiv201015731Z}. The $TS$ value of the source
is 79.5, corresponding to a significance of 8.2$\sigma$ for four free
parameters. Assuming a uniform disk model, we obtain a similar significance with a disk
radius of $0^{\circ}\!.70 \pm 0^{\circ}\!.10$.

To study the significance of the extension of the source,
we define $TS_{\rm ext} = 2 \ln ( {\mathcal L}_{\rm ext}/{\mathcal L}_{\rm ps})$,
i.e., twice the logarithm of the likelihood ratio of an extended source
assumption to a point source assumption \cite{2012nolan_fermi}.
The $TS_{\rm ext}$ for the Gaussian template is about 16.5, which corresponds
to a significance of $\sim 4.1 \sigma$ for an additional free parameter.

\begin{figure}[!htb]
\centering
\includegraphics[width=0.50\textwidth]{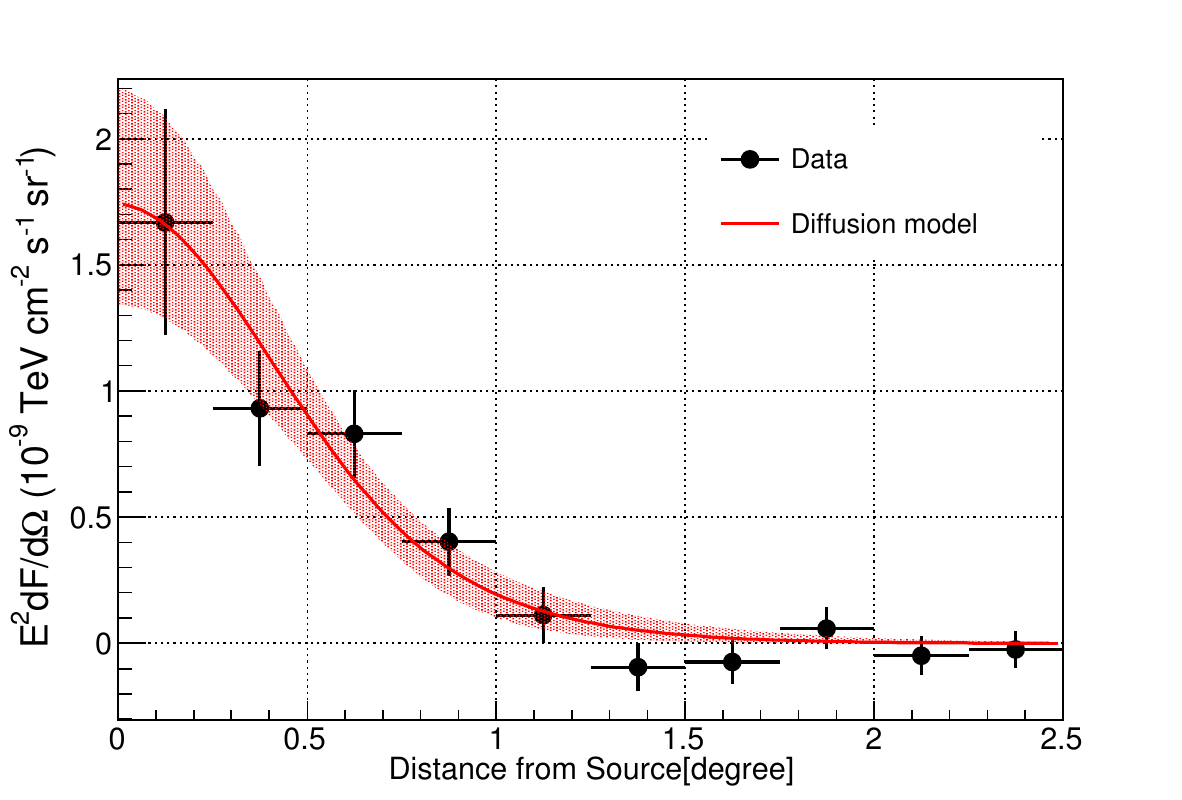}
\caption{One-dimensional distribution of the $>25$~TeV $\gamma$-ray
emission of LHAASO J0621+3755. The solid line and shaded band show the
best fit and $\Delta \chi^2=2.3$ range of the diffusion model fit, which
is the convolution of Eq. (\ref{eq:diffusion}) with the PSF.}
\label{fig:profile}
\end{figure}

To further study the spatial distribution of the source, we use a fitting
form of the morphological distribution from a diffusion model under the
approximation of continuous injection from a point source\footnote{ 
A more detailed modeling of the emission considering e.g., the injection history 
and potential pulsar proper motion \cite{2019MNRAS.484.3491T,2019PhRvD.100l3015D} 
might lead to small differences of the spatial distribution, and hence the 
estimate of the extension parameter.}
\begin{equation}
\label{eq:diffusion}
%f(\theta)\propto\frac{1}{\theta_d(\theta+0.06\theta_d)}\exp(-\theta^2/\theta_d^2),
f(\theta)\propto\frac{1}{\theta_d(\theta+0.085\theta_d)}\exp[-1.54(\theta/\theta_d)^{1.52}],
\end{equation}
to fit the KM2A observed morphology. In the above equation, $\theta$ 
is the angular distance from the source position, and
$\theta_d=\frac{180^\circ}{\pi}\cdot \frac{2\sqrt{D(E_e)t_E}}{d}$
is the typical diffusion extension with $D(E_e)$ as the diffusion
coefficient and $t_E\sim5.5$~kyr as the cooling time of electrons
and positrons with $\sim160$~TeV energies ( see Sec. F of 
{\tt Supplemental Material}~\cite{SM} for the magnetic field and photon
fields used in this work). This formula is a slightly improved 
version of that introduced in Ref.~\cite{2017Sci...358..911A},
and can match the numerical calculation, which includes the diffusion 
of $e^{\pm}$, the inverse Compton scattering (ICS) off the background 
radiation field, and the line-of-sight integral of the $\gamma$-ray 
emission, within a few percent up to a distance as far as $3\theta_d$ 
from the central source (see Sec. G of {\tt Supplemental Material}~\cite{SM}. 
We get the fitted $\theta_d=0^\circ\!.91 \pm {0^\circ\!.20}$ for $E>25$~TeV,
and the TS value of 78.1 for the diffusion model, as given in Table~\ref{table:fit0}. 
 We have tested via MC simulations that the differences among
the three extended templates are not significant (with a largest 
difference of $\sim1.8\sigma$). The one-dimensional distribution
of the number of events after subtracting the estimated background,
together with the fitting $1\sigma$ band of the diffusion model,
are shown in Fig.~\ref{fig:profile}.

\begin{table}[!htb]
\centering
\caption {Results of the morphological analyses of LHAASO J0621+3755.}
\begin{tabular}{ccccccc}
\hline \hline
Template & Extension$^{a}$ ($^{\circ}$) & RA ($^{\circ}$) & Dec ($^{\circ}$) & $TS$ & $N_{\rm p}^{b}$ \\
\hline
Point source & - & $95.56\pm 0.10$ & $37.85\pm 0.07$ &    63.0  & 3 \\
2D Gaussian & $0.40\pm 0.07$ & $95.47 \pm 0.11$ & $37.92 \pm 0.09$  & 79.5 & 4\\
Uniform disk & $0.70\pm 0.10$ & $95.44 \pm 0.11$ & $37.94 \pm 0.09$  & 80.2   & 4 \\
Diffusion & $0.91\pm 0.20$ & $95.48 \pm 0.10$ & $37.90 \pm 0.09$  & 78.1 & 4\\
\hline
\hline
\end{tabular}
\textsuperscript{$a$}\footnotesize{Radius for the uniform disk;
$\sigma$ for the Gaussian model; $\theta_d$ for the diffusion model.}
\textsuperscript{$b$}\footnotesize{$N_{\rm p}$ is the number of parameters in the model.}\\
\label{table:fit0}
\end{table}

\begin{figure}[!htb]
\centering
\includegraphics[width=0.48\textwidth]{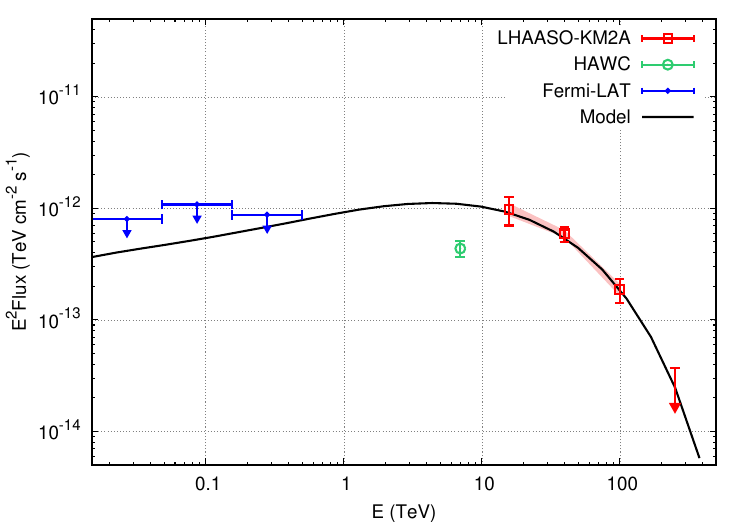}
\caption{The spectrum of LHAASO J0621+3755. The error bars represent statistical
uncertainties, and the shaded band shows the systematic uncertainties.
The HAWC measurement of 3HWC J0621+382 \cite{2020arXiv200708582A} and the
Fermi-LAT $95\%$ upper limits are also shown. The line shows the prediction
based on the pulsar halo model (see Sec. IV).}
\label{fig:J0621_spec}
\end{figure}

Using the Gaussian extension of $0^{\circ}\!.40$, the resulting differential 
flux (TeV$^{-1}$ cm$^{-2}$ s$^{-1}$) , assuming a single power-law form, is
\begin{equation}
\begin{aligned}
\frac{dN}{dE}= ( 3.11 \pm 0.38_{\rm stat} \pm 0.22_{\rm sys} )\times10^{-16}~\\
(E/{40~{\rm TeV}})^{-2.92\pm0.17_{stat}\pm0.02_{sys}}
\end{aligned}
\end{equation}
We derive the fluxes of LHAASO J0621+3755 in four energy bins,
$[10,25]$, $[25-63]$, $[63-158]$, and $[158-398]$ TeV, respectively.
Above 100 TeV, we observed four photon-like events against 0.5
background events, which corresponds to a $3.1\sigma$ statistical
significance. Because the significance in the last energy bin is
smaller than $2\sigma$, a $95\%$ upper limit is derived.
The SED is given in Fig.~\ref{fig:J0621_spec}, where the geometric 
mean energy is used to represent the energy of corresponding bin.
Assuming a power-law with an exponential cut-off improves the
fitting quality very little, which gives a $TS$ value higher by
1.6 but with one more free parameter.

%
%\subsection{Systematic uncertainties}
%The systematic uncertainty in this analysis is very similar to that 
%of the Crab analysis \cite{2020arXiv201006205A}. It mainly comes from 
%the atmospheric model used in the simulation, which affects the {\bf detection 
%efficiency. We estimate this effect by using the variation of event 
%rate during the operating period, which turns out to be $\sim7\%$ on the 
%flux and 0.02 on the spectral index \cite{2020arXiv201006205A}}. 
%When determining the extension of the source, the measurement of the 
%PSF becomes crucial. The systematic uncertainty is estimated to be 
%$\sim 3\%$, by considering the difference in the PSF between simulations 
%and experimental data. In addition, since the Galactic latitude of 
%LHAASO J0621 + 3755 is about $11^{\circ}$, the contribution from 
%Galactic diffuse emission can be neglected.

\section{Multi-wavelength studies}

LHAASO J0621+3755 is a new source in the VHE domain, without a counterpart
in the TeVCat\footnote{http://tevcat.uchicago.edu} \cite{2008ICRC....3.1341W}.
It is potentially associated with 3HWC J0621+382 in the third HAWC catalog
\cite{2020arXiv200708582A}, since the angular distance between two sources is
$0.31^{\circ} \pm 0.32^{\circ}$.
In the GeV energy band, we find two 4FGL sources \cite{2020ApJS..247...33A},
4FGL J0622.2+3749 and 4FGL J0620.3+3804, in the vicinity of LHAASO J0621+3755
(see Fig.~\ref{fig:map25TeV}). 4FGL J0620.3+3804 is associated with the radio
source GB6 J0620+3806 \cite{1996ApJS..103..427G}, and is classified as a ``bcu''
(blazar candidate of uncertain type) in the 4FGL catalog \cite{2020ApJS..247...33A}.
Since LHAASO J0621+3755 shows emission up to 100 TeV energies, we expect that
it should not be associated with 4FGL J0620.3+3804. The other source,
4FGL J0622.2+3749, is a $\gamma$-ray pulsar discovered using the Fermi-LAT
data \cite{2012ApJ...744..105P}. Multi-wavelength counterparts of the
pulsar have been searched for in Ref.~\cite{2012ApJ...744..105P}. No X-ray
or pulsed radio emission has been found. Faint sources and extended 
diffuse emission might exist in the MPIfR surveys of radio continuum emission 
at 408 MHz and 1420 MHz \cite{1982A&AS...47....1H,1982A&AS...48..219R}.
However, after checking the 820 MHz and 4850 MHz images
\cite{1972A&AS....5..263B,1994AJ....107.1829C}, we find no
clear diffuse emission around the pulsar (see Fig.S1
in the {\tt Supplemental Material}~\cite{SM}). The search in the infrared and
optical bands does not reveal counterparts of the pulsar or the
extended halo.

The potential extended GeV $\gamma$-ray emission can give very useful
constraints on the properties of CR injection from the pulsar and
diffusion in the ISM \cite{2019ApJ...878..104X,2019PhRvD.100l3015D,
2020ApJ...889...12Z}. For Geminga, a possible large extended
counterpart of the TeV halo was reported in \cite{2019PhRvD.100l3015D}. 
We therefore analyzed 11.5 years of the Fermi-LAT data to 
search for extended emission associated with LHAASO J0621+3755.
Events of {\tt P8R3} version and {\tt SOURCE} class in a square 
region of $25^\circ\times25^\circ$ with energies between 15 and 500 GeV 
were used. The diffuse models used are {\tt gll\_iem\_v07.fits} and 
{\tt iso\_P8R3\_SOURCE\_V2\_v1.txt}\footnote{http://fermi.gsfc.nasa.gov/ssc/data/access/lat/BackgroundModels.html}.
A binned likelihood method is adopted, with background source model 
XML file generated using 
{\tt make4FGLxml.py}\footnote{http://fermi.gsfc.nasa.gov/ssc/data/analysis/user/}
based on the 4FGL source catalog \cite{2020ApJS..247...33A}.
No clear extended emission has been found. The $95\%$ flux upper limits
have been derived, assuming the predicted spatial template from the 
diffusion model in the relevant energy band. The results are also shown 
in Fig.~\ref{fig:J0621_spec}. For more details of the Fermi-LAT data 
analysis, please refer to the {\tt Supplemental Material}~\cite{SM}.

\section{Interpretation as a pulsar halo}

\begin{table}[!htb]
\centering
\caption {Comparison of the properties of pulsars J0622+3749, 
Geminga, and Monogem.}
\begin{tabular}{ccccccc}
\hline \hline
Name & $P$ & $\dot{P}$ & $L_{\rm sd}$ & $\tau$ & $d$ & Ref.\\
     & (s) & ($10^{-14}$~s~s$^{-1}$) & ($10^{34}$~erg s$^{-1}$) & (kyr) & (kpc) & \\
\hline
J0622+3749 & $0.333$ & $2.542$ & $2.7$ & 207.8 & 1.60 & \cite{2012ApJ...744..105P}\\
Geminga    & $0.237$ & $1.098$ & $3.3$ & 342.0 & 0.25 & \cite{2005AJ....129.1993M}\\
Monogem    & $0.385$ & $5.499$ & $3.8$ & 110.0 & 0.29 & \cite{2005AJ....129.1993M}\\
\hline
\hline
\end{tabular}
\label{table:pulsar}
\end{table}

The multi-wavelength search indicates that LHAASO J0621+3755 is a
VHE $\gamma$-ray-only source possibly associated with PSR J0622+3749.
The pulsar is a middle-aged pulsar, similar to those of Geminga
and Monogem. In Table~\ref{table:pulsar} we compare the main properties
of PSR J0622+3749 with those of Geminga and Monogem. Note
that these three pulsars share similar properties. No clear pulsar
wind nebula (PWN) emission is visible in the radio, infrared,
optical, and X-ray bands. For Geminga and Monogem, weak PWNe have
been shown in the X-ray images
\cite{2006ApJ...643.1146P,2016ApJ...817..129B}.
The lack of PWN emission associated with PSR J0622+3749 can be understood
in terms of its larger distance. Taking the Monogem PWN as an example, the total
unabsorbed $0.5-8$~keV flux of the extended emission (in a small region
with radii from 3$^{\prime\prime}$.5 to 15$^{\prime\prime}$) is $8.3^{+5.7}_{-4.4}\times10^{-15}$~erg~s$^{-1}$~cm$^{-2}$
\cite{2016ApJ...817..129B}. Assuming the same level of PWN emission,
the potential X-ray PWN associated with PSR J0622+3749 would be too
faint given a distance of $1.60/0.29$ times larger.
The multi-wavelength observational results are thus consistent with
the scenario that the VHE $\gamma$-ray emission comes from the
electrons and positrons diffusing out from the pulsar which then
scatter the interstellar radiation field.

%Given the relatively large (pseudo-)distance of PSR J0621+3755,
%it is natural to expect a considerably smaller angular extension
%of LHAASO J0621+3755. The spatial extension of LHAASO J0621+3755 
%is about 11 pc for an angular extension of $0^{\circ}\!.40$, 
%assuming $d=1.6$~kpc.
With the $\theta_d$ values inferred from the morphological fitting,
we obtain a diffusion coefficient $D\approx(8.9_{-3.9}^{+4.5})
\times10^{27}(d/1.6~{\rm kpc})^2$~cm$^2$~s$^{-1}$ for $E_e\sim160$~TeV. 
Here we adopt the approximate scaling relation between electrons and ICS photons
$\bar{E}_e\approx17\bar{E}_{\gamma}^{0.54+0.046\log(\bar{E}_{\gamma}/{\rm TeV})}$
\cite{2017Sci...358..911A}, where $\bar{E}_{\gamma}\approx40$~TeV
is the median energy of photons. The diffusion coefficient is
comparable to the results derived from observations of Geminga and 
Monogem \cite{2017Sci...358..911A}, which is about $4.5\times10^{27}$
cm$^2$~s$^{-1}$ for $E_e\sim100$~TeV, and is significantly smaller than 
that inferred from the CR secondary-to-primary ratios ($\sim 10^{30}-10^{31}$
cm$^2$~s$^{-1}$; \cite{2011ApJ...729..106T,2017PhRvD..95h3007Y}).
Given the current data statistics, the energy dependence of
the diffusion coefficient cannot be robustly determined yet.

We model the wide-band $\gamma$-ray SED of LHAASO J0621+3755 with
an ICS model of $e^{\pm}$ diffusing out from the pulsar or PWN.
A two-zone diffusion model \cite{2018ApJ...863...30F} is assumed,
with a slow-diffusion-coefficient the same as that inferred
from the LHAASO observation, a slow-diffusion region size
of 50 pc, and a diffusion coefficient 300 times larger outside.
The energy-dependence of the diffusion coefficient is assumed to be
proportional to $E^{1/3}$. A super-exponentially cutoff power-law
spectrum, $q(E)\propto E^{-\alpha}\exp[-(E/E_c)^{\beta}]$,
is adopted to describe the injection spectrum of $e^{\pm}$.
For more details of the model setting one can refer to the
{\tt Supplemental Material}~\cite{SM}. The calculated result is shown by 
the solid line in Fig.~\ref{fig:J0621_spec}, with $\alpha=1.5$, 
$\beta=1.0$, $E_c=150$ TeV, and an energy conversion efficiency 
from the pulsar spin-down energy to the accelerated $e^{\pm}$ energy of
$\eta \equiv W_{e^{\pm}}/W_{\rm sd}=40\%\cdot(d/1.6~{\rm kpc})^2$.
Note that the energy budget may also give an independent
support of the slow diffusion scenario of the $e^{\pm}$.
The one-dimensional spatial distribution of the model prediction
is almost identical with the solid line in Fig.~\ref{fig:profile}.
The one-zone slow-diffusion model is found to give too high
fluxes in GeV bands and a very hard spectrum of injected $e^{\pm}$
is required, which seems to be less favored.

Note that the model flux is slightly higher than the flux measured by HAWC
of 3HWC J0621+382 at median energy of $\sim7$ TeV \cite{2020arXiv200708582A}.
The HAWC flux was derived assuming a disk extension of $0^{\circ}\!.5$,
which is smaller than the extension measured by LHAASO ($0^{\circ}\!.70$;
see Table I). We expect that more dedicated analysis of HAWC and
LHAASO-WCDA data will be very helpful in clarifying the spectral behavior
below 10 TeV energies and in better constraining the model parameters.

\section{Conclusions}

Using about ten months of data recorded with the half array of the
LHAASO-KM2A, we discover an extended VHE $\gamma$-ray source, LHAASO
J0621+3755, whose location is consistent with a middle-aged pulsar,
PSR J0622+3749. The source is detected with a significance of 8.2$\sigma$
above 25 TeV, and $3.1\sigma$ above 100 TeV. LHAASO J0621+3755 is
extended with a significance of $\sim4.1\sigma$. A Gaussian fit gives an
extension of $\sim0^{\circ}\!.40\pm0^{\circ}\!.07$. The power-law spectral
index of LHAASO J0621+3755 is ${-2.92 \pm 0.17_{\rm stat} \pm 0.02_{\rm sys}}$.
Together with the flux upper limits from the Fermi-LAT observations,
a downward-curved $\gamma$-ray spectrum is expected. Multi-wavelength 
counterparts of the LHAASO source have been searched for, and no 
associated sources have been found.

The LHAASO and multi-wavelength observations tend to suggest that
LHAASO J0621+3755 is a pulsar halo associated with PSR J0622+3749.
If the VHE $\gamma$-ray emission is interpreted as the ICS emission
from high-energy electrons and positrons that escaped from the pulsar,
the source extension indicates a diffusion coefficient of
$D(160~{\rm TeV})\sim8.9_{-3.9}^{+4.5}
\times10^{27}(d/1.6~{\rm kpc})^2$~cm$^2$~s$^{-1}$.
This is consistent with the slow diffusion scenario of particles in the
turbulent medium around pulsars as inferred from the HAWC observations
of Geminga and Monogem \cite{2017Sci...358..911A}.
The required energy to power the VHE $\gamma$-ray emission is estimated
to be $\sim20\%-40\%$ of the pulsar spin-down energy, assuming a
distance of 1.6 kpc. LHAASO J0621+3755 is thus possibly another 
pulsar halo besides Geminga and Monogem with extensive VHE and 
multi-wavelength studies and the first one with detected $\gamma$-ray 
emission up to 100 TeV energies. The discovery and measured properties 
of LHAASO J0621+3755 may help to establish that the diffusion of CRs 
in the Milky Way is inhomogeneous.

\section{Acknowledgements}

This work is supported in China by the National Key R$\&$D program of China
under the grants 2018YFA0404202, 2018YFA0404201, 2018YFA0404203, 2018YFA0404204,
by the National Natural Science Foundation of China under the grants 11722328,
11851305, 11635011, 11761141001, 11765019, 11775233, U1738205, U1931111, U1931201, 
by Chinese Academy of Sciences, and the Program for Innovative Talents and Entrepreneur 
in Jiangsu, and in Thailand by RTA6280002 from Thailand Science Research and Innovation.
The authors would like to thank all staff members who work at the LHAASO
site which is 4400 meters above sea level year-around to maintain the
detector and keep the electricity power supply and other components of the
experiment operating smoothly. We are grateful to Chengdu Management
Committee of Tianfu New Area for the constant ﬁnancial supports to
research with LHAASO data.

\bibliographystyle{apsrev}
%\begin{thebibliography}
%%\bibliography{refs}

\end{document}